\documentclass[aps,prc,preprint,groupedaddress,superscriptaddress]{revtex4-1}
\usepackage[utf8]{inputenc}
\usepackage{graphicx}
\usepackage{amsmath, amssymb, bm, mathrsfs, gensymb}
\begin{document}
\title{Study of the double Gamow-Teller transitions using the shell model 
approach}
%
%
\author{Naftali Auerbach}
\email[]{auerbach@post.tau.ac.il}
\author{Bui Minh Loc} 
\affiliation{School of Physics and Astronomy, Tel Aviv University, Tel Aviv 
69978, Israel.}

\begin{abstract}
The double Gamow-Teller strength distributions in the lightest 
double beta-decay candidate $^{48}$Ca and its isotope $^{46}$Ca were calculated 
using the nuclear shell model by applying the single Gamow-Teller operator two 
times sequentially on the ground state of parent nucleus. The nuclear 
matrix element of the double Gamow-Teller transition from the ground state to 
the ground state that goes into the double beta decay calculation was shown as 
a small fraction of the total transition.
\end{abstract}
\maketitle
\section{Introduction}
\label{intro}
The double charge-exchange (DCX) processes are a promising tool to study 
nuclear structure in particular nucleon-nucleon correlations in nuclei. In the 
1980s, the DCX reactions using pion beams that were produced in the three meson 
factories at LAMPF, TRIUMF, and SIN were performed \cite{PhysRevLett.60.408, 
PhysRevLett.61.531}.

At present, there is a renewed interest in DCX reactions, to a large extent due 
to the extensive studies of double beta-decay, both the decay in which 
two neutrinos are emitted ($2\nu\beta\beta$) and neutrinoless double beta-decay 
($0\nu\beta\beta$). The pion DCX reactions did not excite the states involving 
the spin, such as the double Gamow-Teller (DGT) state. The DGT strength is  
the essential part of the double beta decay transitions. The pion interacts 
weakly with states involving the spin. It was suggested in the past that one 
could probe such states using DCX reactions with light ions 
\cite{AUERBACH198977, ZHENG1990343}.
The present day, DCX reactions are indeed performed using light ions 
\cite{Cappuzzello2018}. One hopes that such studies might shed some light on the 
nature of the nuclear matrix element of the double beta-decay and serve as a 
“calibration” for the size of this matrix element. These DCX studies might also 
provide new interesting information about nuclear structure. 

One of the outstanding resonances relevant to the double beta-decay is the DGT 
resonance. The notion of a DGT was introduced in Ref. \cite{AUERBACH198977, 
ZHENG1990343}. The DGT strength distributions in even-$A$ Neon isotopes was 
discussed in Ref. \cite{MUTO199213} and recently the calculation for $^{48}$Ca 
was performed in Ref. \cite{PhysRevLett.120.142502}. In both works, the Lanczos 
method \cite{WHITEHEAD1980313} was used. In the present paper, the DGT 
transition strengths in even-$A$ Calcium isotopes are calculated in the full 
$fp$-model space using the nuclear shell model code NuShellX@MSU 
\cite{BROWN2014115, BROWN2001517}. The properties of the DGT distribution are 
examined and limiting cases when the SU(4) holds or when the spin orbit-orbit 
coupling is put to zero are studied. DGT sum rules were derived in Ref. 
\cite{VOGEL1988259, PhysRevC.40.936, MUTO199213, PhysRevC.94.064325}. The DGT 
sum rules in this paper were used as a tool to asses whether in our numerical 
calculations most of the DGT strength is found.

\section{Method of calculation}
\label{sec-1}
The nuclear shell-model wave functions of the initial ground state, 
intermediate states, and final states were obtained from the shell model code 
NuShellX@MSU \cite{BROWN2014115, BROWN2001517} using the FPD6 
\cite{RICHTER1991325} interaction in the complete $fp$-model space. For $J_f 
= 0^+$ in $^{46}$Ti, all 2343 possible states are taken into account. In 
the case of $J_f = 2^+$ in $^{46}$Ti, the calculation was done for 5000 of 9884 
states. We also calculated only 5000 of 14177 $J_f = 0^+$ states in 
$^{48}$Ti. The number of intermediate states is 500 in our work. As one will 
see later, this is enough to exhaust almost the total DGT strength. The number 
of $J = 2^+$ in $^{48}$Ti is too large (61953) to be calculated with the 
present computer codes.

After all wave functions were obtained, the single GT operator was applied two 
times sequentially. First, all transitions from the parent nucleus $0^+$ to all 
$1^+$ intermediate states are calculated and then all transitions from $1^+$ 
intermediate states to each $0^+$ or $2^+$ in the final nucleus are computed.
The single GT operator is defined as
\begin{equation}
 \bm{Y}_{\pm} = \sum_{i = 1}^A \bm{\sigma} t_\pm (i); \quad t_{\pm} = t_x \pm 
it_y,
\end{equation}
with $t_-n = p$ and $t_+ p = n$ where $2t_x$ and $2t_y$ are the Pauli isospin 
operators and $\bm{\sigma}$ is Pauli spin operator. Then the single GT 
transition amplitude $J_i^+ \rightarrow J_f^+$ is
\begin{equation}
 M(GT_{\pm}) = \frac{\langle J_f^+ || \bm{Y}_{\pm} || J_i^+ \rangle}{\sqrt{2J_i 
+ 1}},
\end{equation}
and the GT transition strength given by
\begin{equation}
 B(GT_{\pm}) = |M(GT_{\pm})|^2
\end{equation}
obeys the ``$3(N-Z)$'' sum rule. 

The dimensionless DGT transition amplitude is defined as
\begin{equation}
 M(DGT_{\pm}) (J_f) = \sum_n M(GT_{\pm}; i\rightarrow n) M(GT_{\pm}; 
n\rightarrow f),
\end{equation}
where $n$ are the intermediate states. Note that this is a coherent sum. 
The DGT strength is given by
\begin{equation}
 B(DGT_{\pm}) (J_f) = |M(DGT_{\pm}) (J_f)|^2.
\end{equation}
The DGT sum rules for $J_f = 0^+$ and $J_f = 2^+$ are given in Ref. 
\cite{VOGEL1988259, PhysRevC.40.936, MUTO199213, PhysRevC.94.064325}:
\begin{eqnarray} \label{DGTsumrule}
 S_{\text{DGT}}^{J_f = 0} =  6(N-Z)(N-Z+1) - 2\Delta, \nonumber \\
 S_{\text{DGT}}^{J_f = 2} = 30(N-Z)(N-Z-2) + 5\Delta,
\end{eqnarray}
where
$ \Delta = \sqrt{2} \langle 0| [\bm{Y}_+ \times \bm{Y}_-]^{(1)} \cdot 
\bm{\varSigma} - \bm{\varSigma} \cdot [\bm{Y}_- \times \bm{Y}_+]^{(1)} | 
0\rangle,$ with $\bm{\varSigma} = \sum_i \bm \sigma (i)$.
There is factor of three difference between the equations in 
Ref.~\cite{MUTO199213, PhysRevC.94.064325} and our work because the spin 
operator is not projected.

\section{Results and discussions}
\label{sec-2}
We present here only the results for the two heaviest Calcium isotopes, 
$^{46}$Ca and $^{48}$Ca. Note that the double beta-decay from the ground state 
of $^{48}$Ca to the ground state of $^{48}$Ti is energetically allowed and 
studied extensively. A review was given in Ref.~\cite{Engel17review}.

\begin{figure}
\includegraphics[scale=0.5]{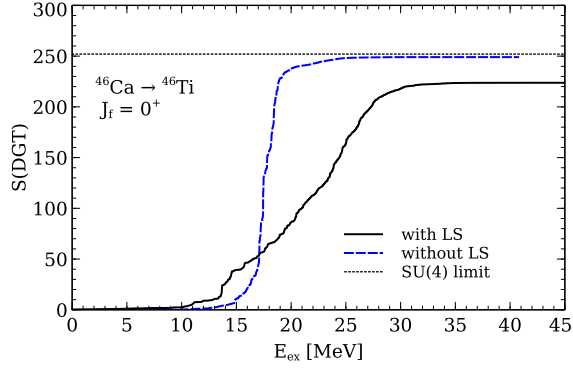}
\caption{The cumulative sum of the DGT strength $B(\rm{DGT}; 
0^+ \rightarrow 0^+)$ in $^{46}$Ca. \label{DGTCa46cumsum0}}
\end{figure}

\begin{figure}
\includegraphics[scale=0.5]{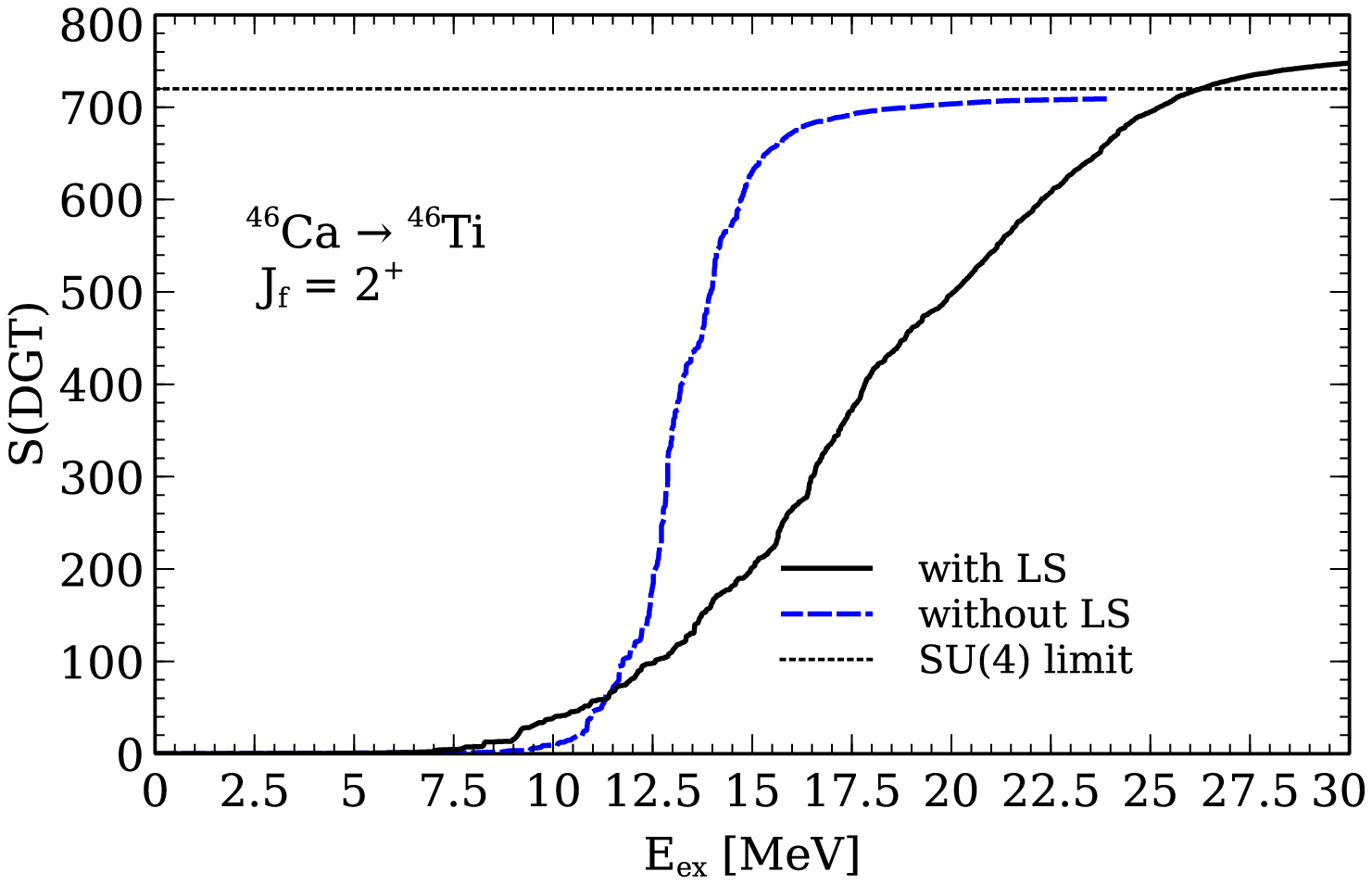}
\caption{The cumulative sum of the DGT strength $B(\rm{DGT}; 
0^+ \rightarrow 2^+)$ in $^{46}$Ca.\label{DGTCa46cumsum2}}
\end{figure}

\begin{figure}
\includegraphics[scale=0.5]{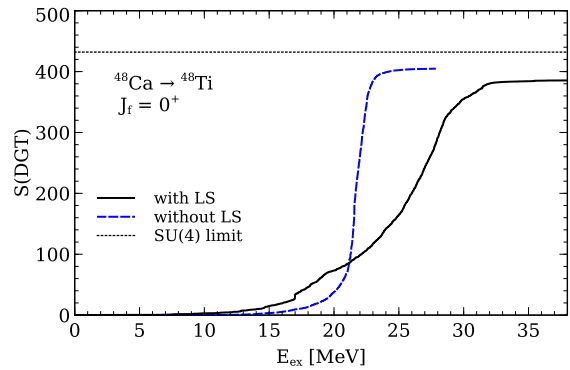}
\caption{The cumulative sum of the double Gamow-Teller strength $B(\rm{DGT}; 
0^+ \rightarrow 0^+)$ in $^{48}$Ca.\label{DGTCa48cumsum0}}
\end{figure}

\begin{table}
\centering
\caption{The total strength of DGT transition.}
\label{tab-1}
\begin{tabular}{|l|r|r|r|r|r|}
\hline
Nucleus & $^{46}$Ti $0^+$ & $^{46}$Ti 
$2^+$ & $^{48}$Ti $0^+$ \\
\hline
Calculation & 223.7 & 752.6 & 385.0 \\
Sum rule & $\leq 252$ & $\geq 720$ & $\leq 432$ \\ 
$0^+_{1}$ to $J^+_{1}$
& 0.201 & 0.017 & 0.109 \\ 
\hline
\end{tabular}
\end{table}
To ensure we can exhaust all the DGT strength, the sum rules of the DGT 
operator are presented numerically and compared to the values given in 
Ref.~\cite{VOGEL1988259, PhysRevC.40.936, PhysRevC.94.064325}. Our 
results are given in Table \ref{tab-1} and they are in agreement with the 
results in Ref.~\cite{PhysRevC.40.936}  for $J^{\pi}_f = 0^+$ and the recent 
work of Ref.~\cite{PhysRevC.94.064325} for both $J^{\pi}_f = 0^+$ and $2^+$.
After the entire distributions are obtained, the cumulative sums of the DGT 
transitions are shown in Fig.~\ref{DGTCa46cumsum0}--\ref{DGTCa48cumsum0}. We 
remind that the entire DGT distributions of even-$A$ Ne isotopes were obtained 
in Ref.~\cite{MUTO199213} but a different method of calculation from our 
work was used. Note that Ref.~\cite{PhysRevC.40.936, PhysRevC.94.064325} 
calculated the DGT sum rule indirectly and therefore gave only the value of the 
total sum, not the cumulative sum. In 
Fig.~\ref{DGTCa46cumsum0}--\ref{DGTCa48cumsum0}, 
the horizontal line represents the value of the DGT strength in 
the case when the SU(4) is a good symmetry. It is the upper limit for DGT sum 
rule of $J_f = 0^+$ and lower limit for $J_f = 2^+$. We also show in above 
figures the computed strength in the limiting case when the spin-orbit coupling 
is put to zero (the SU(4) symmetry is approximately restored) 
\cite{PhysRevC.96.044319}. Because all possible $J_f = 0^+$ final states in 
$^{46}$Ti were taken into account, Fig.~\ref{DGTCa46cumsum0} shows that the sum 
rule, in this case, was exhausted and when the spin-orbit coupling is put to 
zero its cumulative sum approach the limit value (the horizontal line). 
Fig.~\ref{DGTCa46cumsum2} and Fig.~\ref{DGTCa48cumsum0} show that the 
cumulative sums are still increasing because the calculations were limited up to 
5000 final states.

\begin{figure}[h]
\centering
\includegraphics[scale=0.5]{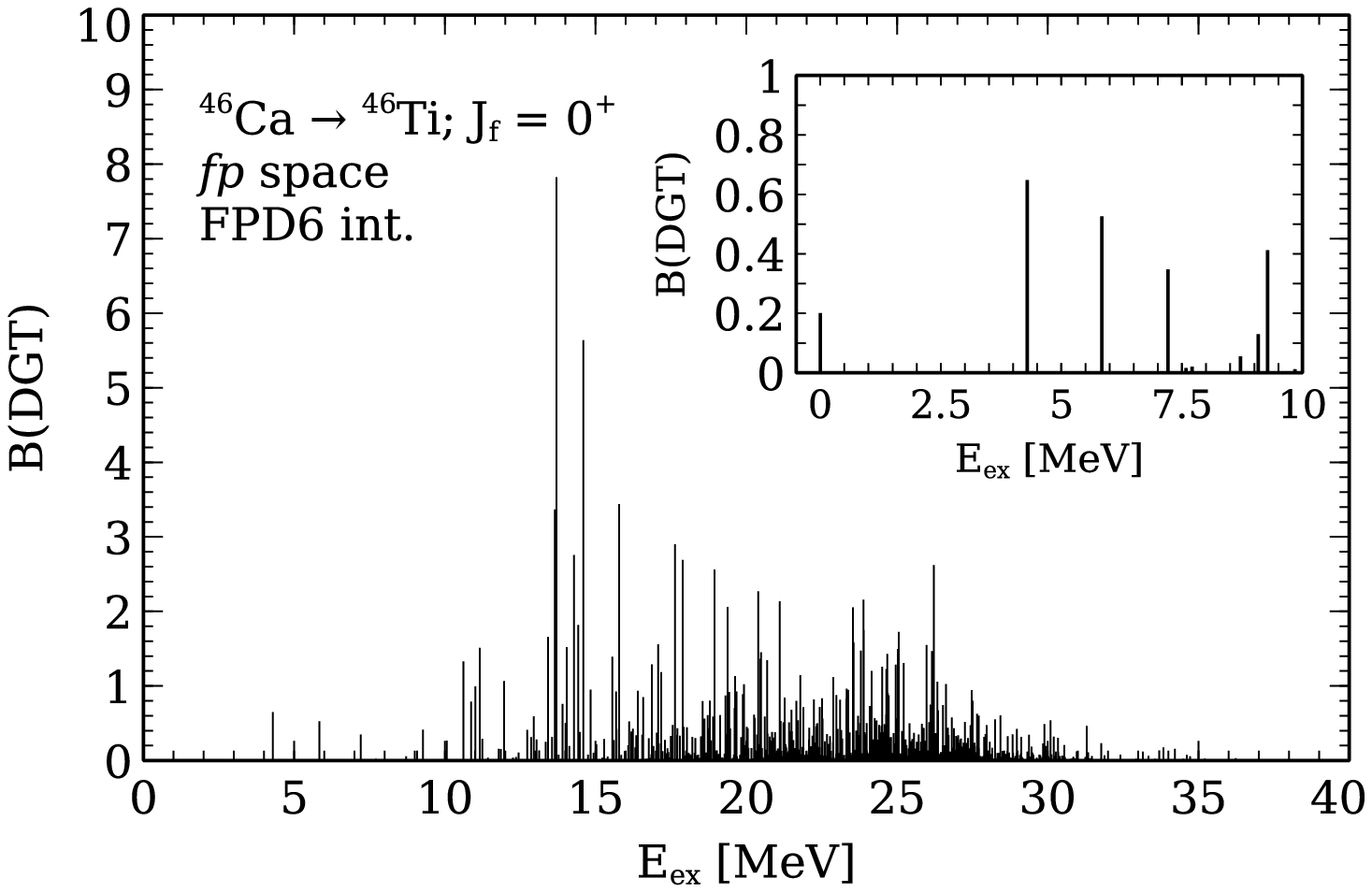}
\caption{$B(\rm{DGT}; 0^+ \rightarrow 0^+)$ for $^{46}$Ca. The insert shows the 
DGT strength in the low-lying states. 
\label{DGTCa461}}
\end{figure}

\begin{figure}[h]
\centering
\includegraphics[scale=0.5]{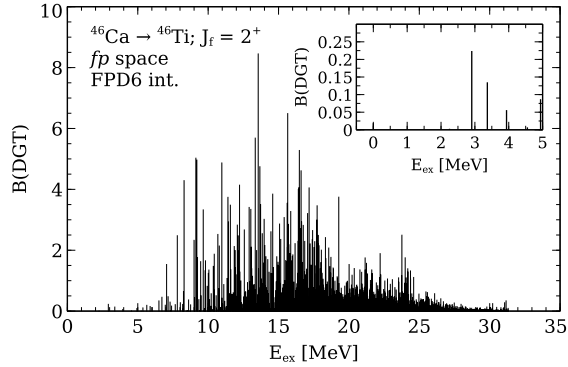}
\caption{The same as Fig~.\ref{DGTCa461} but now for $B(\rm{DGT}; 0^+ 
\rightarrow 2^+)$.
\label{DGTCa462}}
\end{figure}

\begin{figure}[h]
\centering
\includegraphics[scale=0.5]{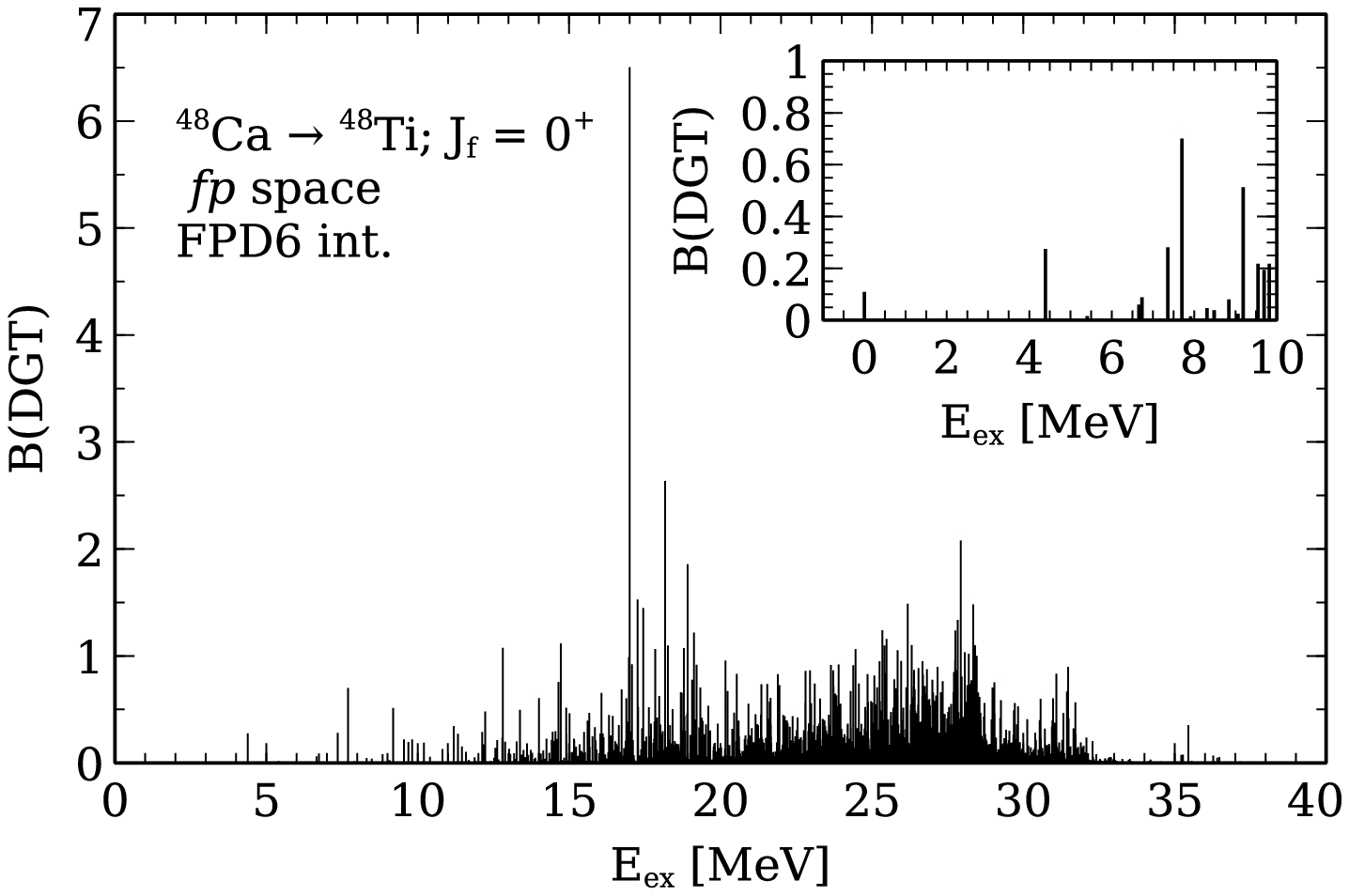}
\caption{The same as Fig~.\ref{DGTCa461} but now for $^{48}$Ca.
\label{DGTCa482}}
\end{figure}

The detailed DGT strength distributions are shown in Fig.~\ref{DGTCa461}, 
Fig.~\ref{DGTCa462} for $^{46}$Ca, and in Fig.~\ref{DGTCa482} for $^{48}$Ca. 
Fig.~\ref{DGTCa461}--Fig.~\ref{DGTCa482} contain inserts which show the DGT 
strength in the low-lying states of $^{46,48}$Ti. The transition strength is a 
very tiny fraction of the total strength. For example, the strength in the 
ground state of $^{48}$Ti is only $3\times10^{-4}$ of the total strength (see 
Table \ref{tab-1}). This strength enters in the calculation of the double 
beta-decay.

\begin{figure}[h]
\centering
\includegraphics[scale=0.5]{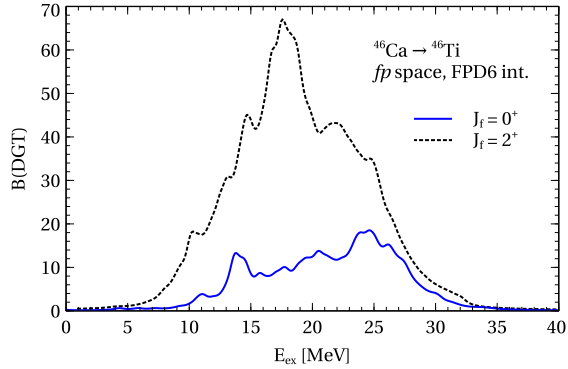}
\caption{$B(\rm{DGT}; 0^+ \rightarrow 0^+; 2^+)$ for $^{46}$Ca. The strengths 
are smoothed by using Lorentzian averaging with the width of 1 MeV.
\label{DGTCa467}}
\end{figure}

\begin{figure}[h]
\centering
\includegraphics[scale=0.5]{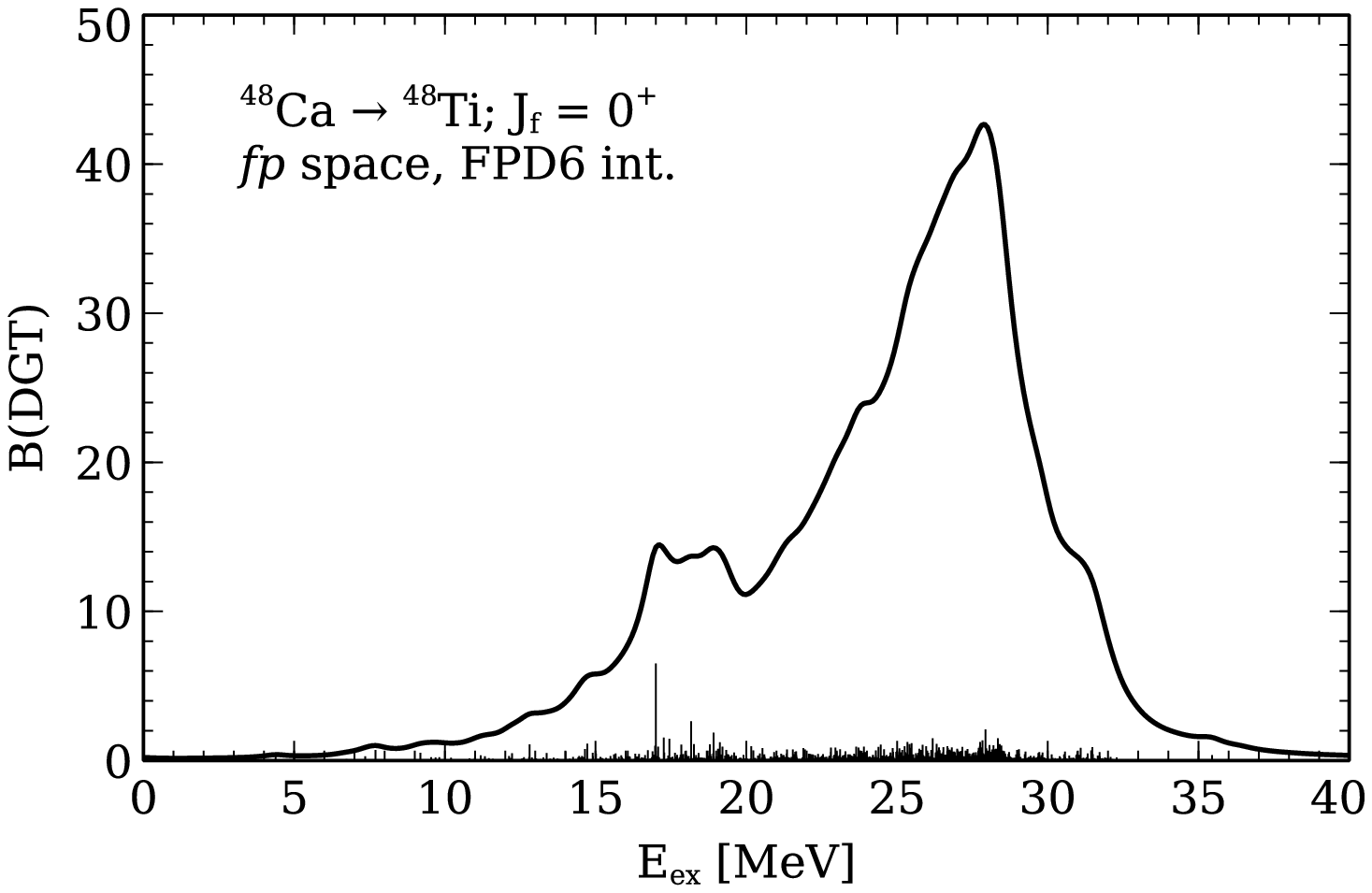}
\caption{$B(\rm{DGT}; 0^+ \rightarrow 0^+)$ for $^{48}$Ca. The strengths 
are smoothed by using Lorentzian averaging with the width of 1 MeV.
\label{DGTCa481}}
\end{figure}

After that, all the strengths are spread by using Lorentzian averaging with 
the width of 1 MeV. Fig.~\ref{DGTCa467} shows that the DGT transition to the 
$J_f = 2^+$ is stronger than the transition to $J_f = 0^+$. 
Fig.~\ref{DGTCa481} shows the distribution in $^{48}$Ca before and after the 
Lorentzian averaging. We observe that the distributions are not single-peaked. 
There are at least two peaks and in some nuclei as many as four major 
peaks. We should remind that the single GT resonances have at least two peaks 
\cite{GOODMAN1982241}.

The average energy of the DGT strength $\overline{E}$ is defined as:
\begin{equation}
 \overline{E} = \frac{\sum_f E_f B_f(DGT_{-})}{\sum_f B_f(DGT_{-})},
\end{equation} 
where $B_f(DGT_{-})$ is the DGT transition at the energy $E_f$.
In $^{46}$Ti, this energy for the $J = 0^+$ is $\overline{E} = 21.2$ MeV and 
for the $J = 2^+$ it is lower $\overline{E} = 18.0$ MeV. In $^{48}$Ti we 
calculated only the $J = 0^+$ DGT distribution. Its average energy is 
$\overline{E} = 24.6$ MeV.
In a recent paper \cite{Ejiri17}, the experimental results for the DCX reaction 
$^{56}$Fe($^{11}$B, $^{11}$Li) are presented. In 
this reaction several resonances were excited. There is a peak at 25 MeV 
excitation, that the authors indicate that it could be the DGT resonance.

\section{Conclusion}
The DCX interaction involving ions is much more complicated than the DGT 
operator, and the reaction mechanism is more evolved than the simple sequential 
process. However the DCX reaction will excite the DGT strength, and when the 
energy of the projectile is high enough it will excite the DGT resonance, as 
well as low-energy states containing DGT strength. A comparison between theory 
and the experimental cross-sections will provide useful information about the 
DGT strength and thus help to learn more about the double beta-decay nuclear 
matrix element. More work is needed on the DCX reaction theory before this goal 
is achieved.

\section*{Acknowledgements}
We wish to thank B. A. Brown, Chavdar Stoyanov, and Vladimir Zelevinsky for 
discussions. This project was supported by the Bulgarian and Israeli Academies 
and by the US-Israel Binational Science Foundation (2014.24).

\end{document}